# Fiber Bragg grating-based acoustic sensing system enabled by ML-trained, sub-picometer-tunable hybrid III-V/SiN lasers

**Prabhav Gaur[1], Premanand Chandramani[1], Yossef Ehrlichman[1], Mohammed Alshamari[2], Yufei Chu[2], Abu Mitul[2], John Simons[1], Ibrahim G. Yayla[1], Ming Han[2], and Ashok V. Krishnamoorthy[1]**

*[1] Axalume Inc., 4040 Sorrento Valley Rd., Ste A, San Diego, CA 92121, USA*

*[2]Michigan State Univ., ECE Dept. 428 S. Shaw Lane, East Lansing, USA*

*Author e-mail address: ashok@axalume.com*

**Abstract:** Distributed acoustic emission (AE) sensing is critical for early detection of structural degradation, yet conventional electrical sensors are difficult to scale and fiber-based approaches are limited by interrogation complexity and resolution. Here, we report an intelligent fiber Bragg grating (FBG) sensing system enabled by machine learning (ML)-trained hybrid III–V/SiN tunable lasers that achieve uniform, mode-hop-free, sub-picometer wavelength control. A supervised gradient-descent algorithm is used to learn the nonlinear electro-thermal tuning space of Vernier-based external-cavity lasers, enabling continuous tuning with <0.1 pm resolution and <0.5 dB power variation. This capability allows precise alignment to FBG reflection slopes for high-sensitivity acoustic detection. We demonstrate a four-laser interrogation system monitoring 16 FBG sensors distributed across multiple metallic structures, operating over a 35 nm wavelength span. The system autonomously identifies sensor resonances, dynamically tracks spectral shifts, and reconfigures interrogation wavelengths in response to localized acoustic events. Using pencil-lead break tests as calibrated AE sources, we show simultaneous multi-channel detection and adaptive spatial localization. The combination of narrow linewidth (<10 kHz), wide tunability, and ML-driven calibration enables robust, scalable, and high-resolution sensing. This approach establishes a pathway toward fully autonomous, distributed photonic sensing networks for real-time structural health monitoring.

## Introduction

The ability to detect and localize early-stage structural degradation is critical for ensuring the safety and longevity of infrastructure, aerospace systems, and industrial assets. Acoustic emission (AE) sensing has emerged as a powerful technique for monitoring dynamic processes such as micro-cracking, fatigue, delamination, and impact damage, as these phenomena generate transient elastic waves that can be detected in real time [1]. Conventional AE sensing systems rely on distributed electrical sensors, which require extensive wiring, are susceptible to electromagnetic interference, and are difficult to scale to large or complex structures [2]. These limitations motivate the development of optical sensing approaches that can provide distributed, high-resolution, and robust monitoring capabilities.

Fiber Bragg gratings (FBGs) offer an attractive solution for quasi-distributed sensing, as multiple sensors can be multiplexed along a single optical fiber and interrogated remotely with high sensitivity [3]. FBGs encode strain and acoustic perturbations as shifts in their Bragg wavelength, enabling precise measurement without electrical components at the sensing site [4]. However, the performance of FBG-based sensing systems is fundamentally limited by the capabilities of the interrogation source. High-resolution, continuous, and stable wavelength tuning is required to accurately track small spectral shifts, particularly for high-frequency acoustic signals [5]. Existing interrogation techniques often rely on broadband sources, swept lasers, or interferometric methods, which can suffer from limited resolution, complex system architectures, or reduced robustness under environmental variations [6] [7] [8] [9].

External-cavity tunable lasers based on hybrid III–V and silicon photonics platforms have emerged as promising candidates for compact, narrow-linewidth, and widely tunable light sources [10] [11] [12] [13]. In particular, Vernier ring-based architectures enable large tuning ranges with high spectral selectivity [14] [15]. However, achieving continuous, mode-hop-free tuning with sub-picometer resolution remains a significant challenge. The relationship between electrical tuning parameters and lasing wavelength is highly nonlinear and device-specific, influenced by fabrication tolerances, thermal crosstalk, and environmental drift. As a result, conventional open-loop control or pre-calibrated lookup tables are insufficient for reliable, high-precision operation.

Recent advances in machine learning (ML) [16] [17] [18] provide a new pathway for overcoming these limitations by enabling adaptive, data-driven control of complex photonic systems. By learning the underlying mapping between control parameters and optical response, ML algorithms can dynamically optimize laser operation in the presence of nonlinearities and uncertainties. Integrating such intelligence directly into the tuning process has the potential to transform tunable lasers into self-calibrating, robust sources suitable for demanding sensing applications.

Here, we report an intelligent FBG-based acoustic sensing system enabled by ML-trained hybrid III–V/SiN external-cavity tunable lasers, which we call ISET (Intelligent Si External-cavity Tunable) lasers. We implement a supervised learning framework that achieves continuous, mode-hop-free wavelength tuning with sub-picometer resolution and low power variation, enabling precise alignment to FBG reflection features. Using an array of such lasers, we demonstrate a distributed sensing architecture capable of simultaneously interrogating multiple FBG sensors and dynamically reconfiguring in response to acoustic events. Controlled pencil-lead break (PLB) experiments validate the ability of the system to detect and track transient AE signals across multiple structures. This work establishes a scalable and robust approach to integrating intelligent photonic sources with distributed sensing networks, opening new opportunities for real-time structural health monitoring.

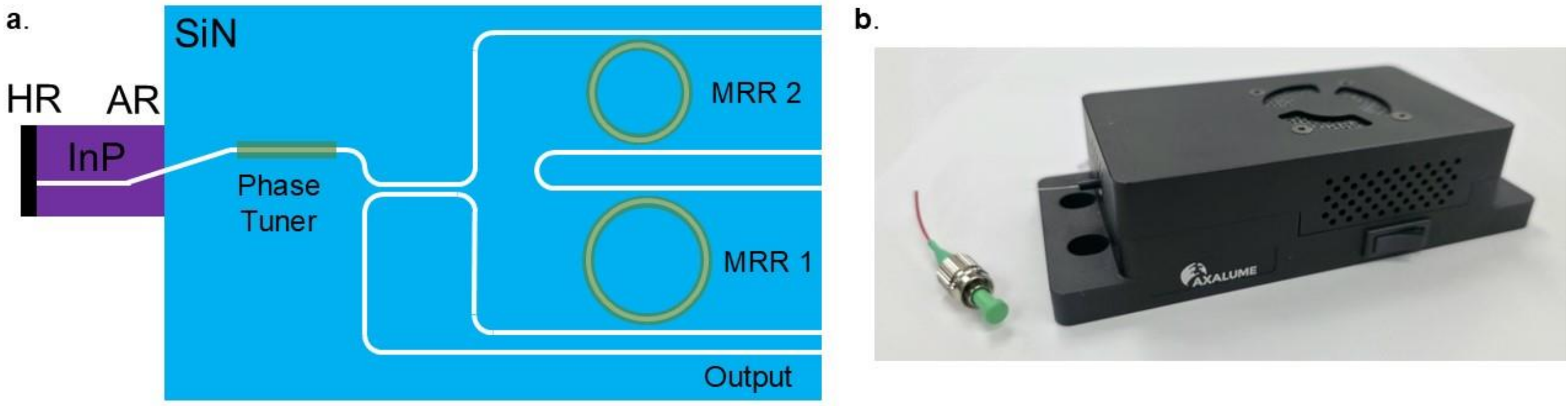

**Fig. 1 | Hybrid InP-Si/SiN Vernier ring external-cavity tunable laser and packaging. a** Schematic of a hybrid external-cavity laser combining an InP gain section with a Si-on-insulator or SiN-on-insulator photonic circuit. A Vernier configuration using dual microring resonators (MRR1 and MRR2), along with integrated phase and thermal tuners, enables precise selection and continuous tuning of the lasing wavelength. **b** Butterfly-packaged Intelligent Si External-cavity Tunable (ISET) laser module integrated with an electronic controller, providing compact, stable, and user-controlled wavelength tuning for practical deployment.

# Results

## Hybrid InP-Si/SiN Vernier ring external-cavity tunable laser

Fig. 1a illustrates the architecture of the hybrid III–V/SiN external-cavity tunable laser used in this work. An InP-based reflective semiconductor optical amplifier (RSOA), with high-reflection (HR) and anti-reflection (AR) facets, is edge-coupled to a low-loss silicon nitride (SiN) photonic integrated circuit. The SiN circuit incorporates two thermally tunable microring resonators (MRR1 and MRR2) arranged in a Vernier configuration to provide coarse wavelength selectivity across a wide tuning range. A phase tuner integrated in the bus waveguide enables fine, continuous wavelength adjustment and ensures alignment of the external cavity modes. The combined action of the Vernier filters and phase shifter enables narrow-linewidth (<10 kHz), mode-hop-free tuning with sub-picometer resolution. The laser output is extracted from the SiN waveguide, forming a compact, high-coherence, and widely tunable source.

Fig. 1b shows the packaged implementation of the device, where the hybrid laser is integrated into a compact module with fiber coupling and electrical interfaces for heater and gain control. The module supports precise electronic tuning and is designed for scalable deployment in multi-laser interrogation systems. In this work, multiple such devices are operated in parallel and controlled via machine-learning-assisted calibration to enable autonomous, high-resolution interrogation of distributed fiber Bragg grating sensors for acoustic emission sensing.

Coarse wavelength tuning in hybrid III–V/SiN external-cavity lasers is typically achieved through a low-resolution grid search over the thermal tuning parameters of the Vernier microring resonators as shown in Fig. 2a. By sweeping the heater currents applied to the rings (and optionally the phase tuner), discrete lasing states are identified across the gain bandwidth. This process enables approximate mapping between control parameters and lasing wavelength, providing initial operating points for each spectral region. However, coarse tuning inherently forces the laser to transition between longitudinal cavity modes, resulting in mode hopping, discontinuous wavelength steps, and power fluctuations.

Achieving fine, continuous tuning is significantly more challenging due to the complex, nonlinear relationship between heater currents, cavity modes, and filter resonances. Fabrication-induced variations, thermal crosstalk between heaters, and temperature-dependent changes in refractive index further distort this mapping, making it non-unique and device-specific. Small perturbations in control parameters can lead to abrupt mode transitions rather than smooth wavelength shifts. Additionally, maintaining simultaneous alignment of multiple resonant elements (e.g., dual rings and phase section) is required to preserve single-mode operation with stable output power. Fig. 2b and Fig. 2c show such an example, where tuning the laser current and using thermos-optic effect on the entire device is unable to provide continuous fine tuning. The wavelength sweep is distorted by cavity mode hop, vernier ring mode hop or misalignment of ring modes and cavity modes. As a result, conventional open-loop or lookup-table-based approaches are insufficient for sub-picometer, mode-hop-free tuning, necessitating adaptive optimization techniques such as supervised machine learning.

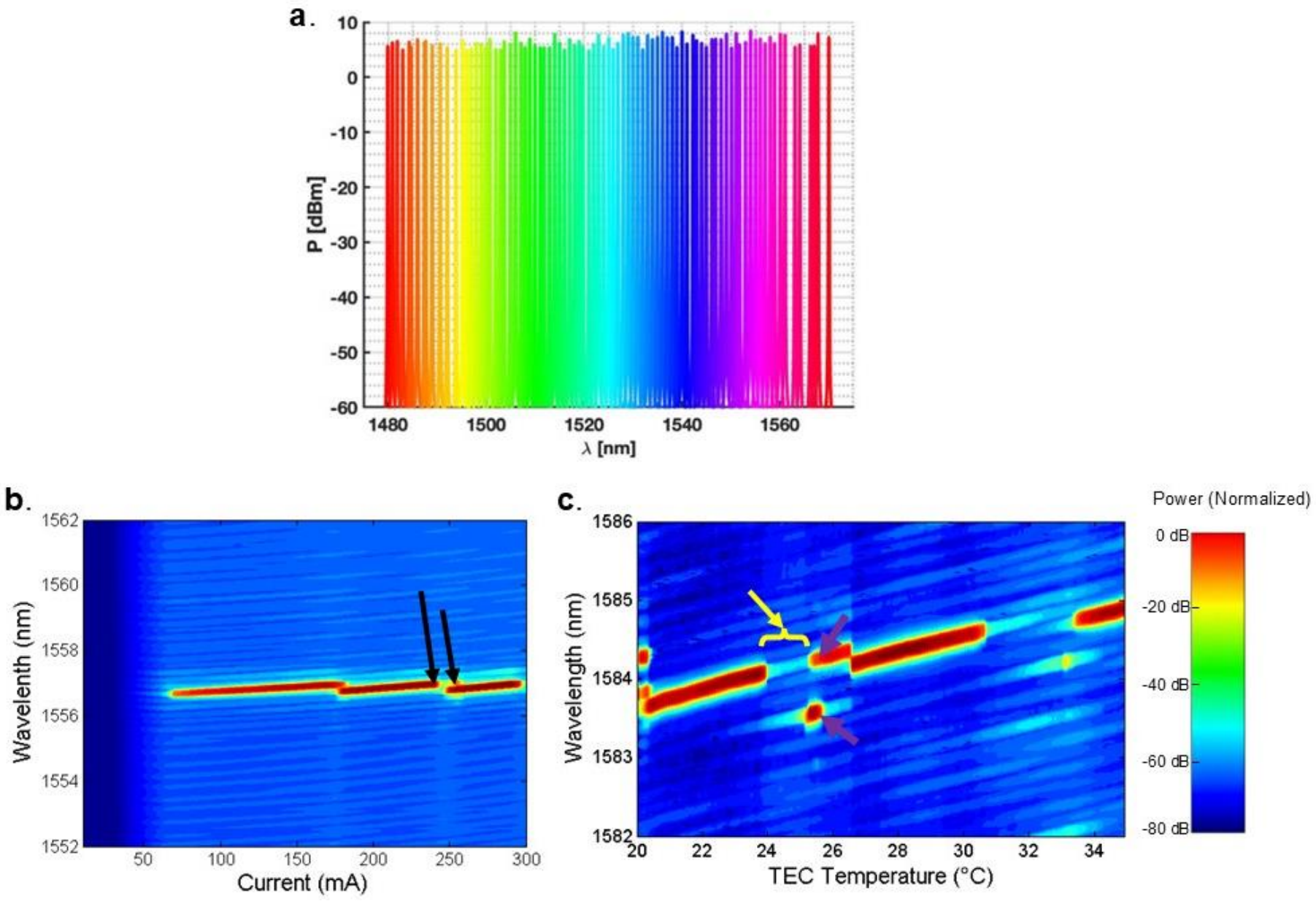

**Fig. 2 | Coarse tuning and mode-hopping behavior in hybrid III–V/SiN external-cavity lasers. a** Superimposed optical spectra obtained during coarse tuning via grid search of the thermal tuning parameters and mode-hopping of vernier ring pair, showing discrete lasing states across the gain bandwidth. The discontinuous spectral evolution reflects mode-hopping between adjacent vernier modes, limiting tuning resolution and stability. **b** Wavelength evolution as a function of laser bias current, highlighting regions of abrupt wavelength transitions (black arrows) associated with mode-hop to neighboring cavity. These discontinuities arise from misalignment between cavity modes and filter resonances during current-induced refractive index changes. **c** Wavelength evolution with temperature (TEC control), illustrating thermally induced red-shift of cavity modes and intermittent mode hops (highlighted), driven by the thermo-optic effect and resonance detuning. The wavelength might be missing for a particular band (yellow arrow) or going through larger mode-hop in vernier pair (purple arrow). Together, these results demonstrate the challenges of achieving continuous, mode-hop-free tuning using conventional open-loop control, motivating the need for machine learning–based optimization of tuning parameters.

## Supervised machine learning based fine tuning

To overcome the limitations of coarse tuning and enable continuous, mode-hop-free operation, we implement a supervised machine learning (ML) algorithm that learns the nonlinear mapping between electrical tuning parameters and the resulting optical response. The schematic of the setup is shown in Fig. 3a. The tuning space, defined by the heater currents applied to the two Vernier microring resonators and the phase section, is inherently multidimensional and non-convex due to fabrication variations, thermal crosstalk, and device-specific nonlinearities. As a result, analytical modeling or static lookup tables are insufficient for achieving precise control.

The ML algorithm is initialized using coarse grid-search data to identify approximate operating points near the desired wavelength. Starting from these points, a gradient-descent optimization is performed using optical power as the cost function, ensuring alignment of the cavity modes with the filter resonances. The ring heater currents are iteratively perturbed to estimate local gradients, while the phase tuner is incremented to shift the lasing wavelength. The laser current and TEC temperature, are fixed during the entire process. At each step, the algorithm re-optimizes the resonator alignment, maintaining single-mode operation and suppressing mode hopping. The resulting tuning parameters are stored in a lookup table, which can be further interpolated to enable continuous wavelength control across the tuning range.

Using this approach, the laser achieves stable, mode-hop-free tuning across multiple wavelength regions spanning the C-band. As shown in Fig. 3b, superimposed spectra demonstrate smooth and continuous wavelength shifts with consistent lineshape and minimal power variation (<0.5 dB). The ability to reproducibly tune and stabilize the laser across distinct spectral regions highlights the robustness of the ML framework in compensating for device nonlinearities and environmental variations. The fine-tuning capability is further demonstrated through high-resolution wavelength sweeps, as shown in Fig. 3c. The laser exhibits continuous tuning with sub-picometer resolution (<0.1 pm) over a narrow spectral window, without observable mode hops or spectral discontinuities. The overlapping traces confirm excellent repeatability and stability, enabled by precise control of the cavity alignment at each tuning step. This level of resolution and stability is critical for applications such as fiber Bragg grating sensing, where accurate tracking of small wavelength shifts is required.

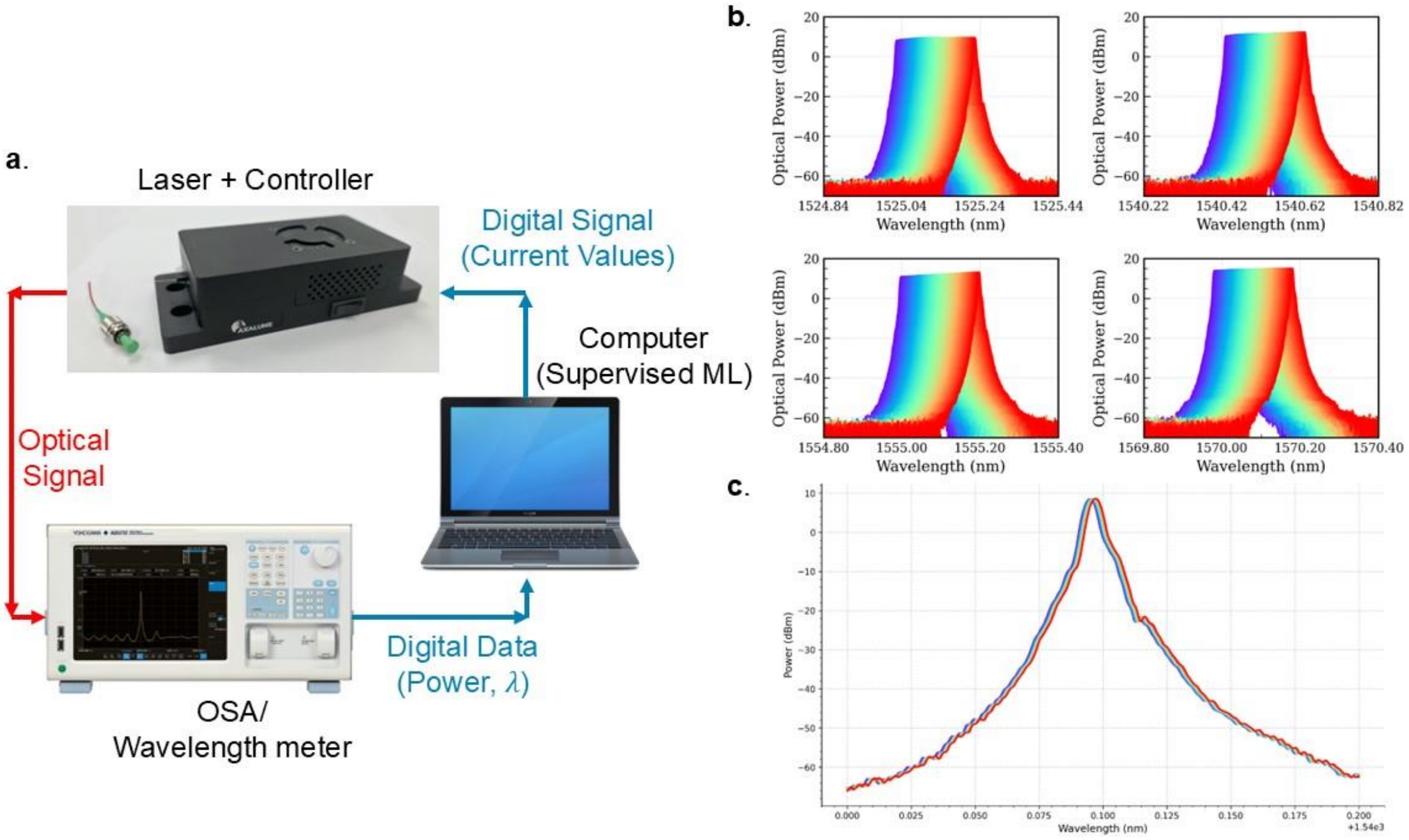

**Fig. 3 | Supervised machine learning–enabled fine tuning and spectral stabilization of hybrid III–V/SiN lasers. a** Closed-loop experimental setup for ML-based laser tuning. The optical output from the laser is measured using an optical spectrum analyzer (OSA) and fed to a computer implementing a supervised learning algorithm. The algorithm processes wavelength and power data and updates the heater currents (digital control signals) to iteratively optimize laser operation. **b** Superimposed optical spectra at multiple wavelengths demonstrating continuous, mode-hop-free tuning across different regions of the gain bandwidth. The smooth spectral evolution with consistent lineshape and minimal power variation indicates successful fine tuning and stabilization. **c** High-resolution (<0.1 pm) wavelength sweep showing continuous, mode-hop-free tuning over a narrow spectral region. The overlapping traces demonstrate precise control of the lasing wavelength with excellent repeatability and low power variation, highlighting the capability of ML-based optimization for sub-picometer resolution tuning.

## Distributed acoustic sensing system

The ML-trained tunable lasers are integrated into a multi-channel interrogation system for distributed AE sensing using FBGs, as illustrated in Fig. 4. A set of narrow-linewidth, widely tunable lasers are multiplexed to simultaneously interrogate multiple reflective FBG sensors inscribed along a single optical fiber. Each FBG acts as a wavelength-selective reflector, and the lasers are dynamically positioned on the slope of the corresponding reflection notch to maximize sensitivity to strain-induced wavelength shifts. Transient acoustic events modulate the effective Bragg wavelength, resulting in intensity variations at the photodetector that are proportional to the incident acoustic signal.

The experimental setup consists of four intelligent silicon external-cavity tunable (ISET) lasers combined through a 4×1 coupler and routed via an optical circulator to a network of 16 FBGs distributed across multiple aluminum plates of different size and forms. The reflected signals are separated using wavelength-selective filters and detected using photodetectors followed by transimpedance amplification and data acquisition. A microprocessor implements real-time control and signal processing, including separation of low-frequency components (temperature and quasi-static strain) from high-frequency acoustic signals.

To demonstrate acoustic sensing capability, pencil-lead break [19] tests are performed as calibrated AE sources on the instrumented plates. Upon detection of an AE event at any sensor location, the system dynamically reallocates the interrogation wavelengths, enabling all lasers to focus on FBGs in the vicinity of the active region. This adaptive sensing strategy allows simultaneous multi-point detection and improves spatial localization of acoustic events. The combination of ML-enabled wavelength control and distributed FBG sensing enables continuous, high-resolution monitoring of structural dynamics with high sensitivity and robustness.

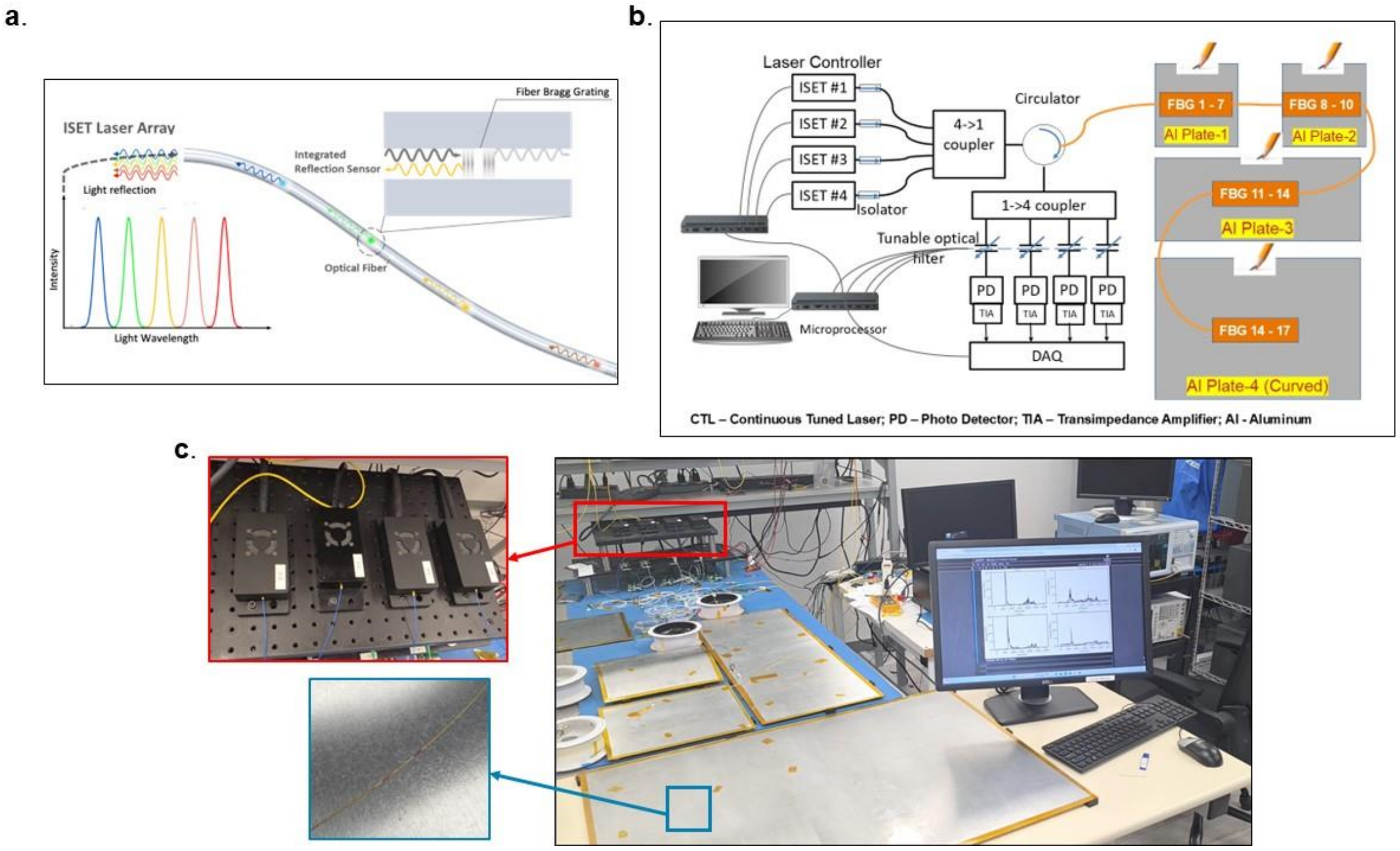

**Fig. 4 | Fiber Bragg grating (FBG) based distributed acoustic sensing system enabled by ML-trained ISET lasers. a** Operating principle of the FBG sensing system. A multi-wavelength laser array interrogates reflective FBG sensors embedded along a single optical fiber. Each FBG reflects a specific wavelength, enabling multiplexed sensing of strain or acoustic perturbations through wavelength shifts. **b** Schematic of the experimental setup. Four ML-trained intelligent silicon external-cavity tunable (ISET) lasers are combined via a 4×1 coupler and routed through a circulator to interrogate multiple FBGs distributed across different aluminum plates. Reflected signals are demultiplexed using tunable optical filters and detected by photodetectors, with signals processed via transimpedance amplifiers and data acquisition (DAQ). The system dynamically identifies and tracks FBG resonances and reconfigures interrogation wavelengths in response to acoustic events.
**c** Photograph of the experimental setup, showing the multi-laser array, fiber network, instrument rack, and aluminum test plates instrumented with FBG sensors. Insets highlight the packaged laser modules and the fiber with inscribed gratings used for distributed sensing.

The sensing mechanism relies on accurate identification and tracking of the reflection notch of each FBG, an example of which is shown in Fig. 5a. FBG spectra at different wavelengths is shown in SI-I. The FBGs exhibit a narrowband reflection dip within the broadband optical spectrum, corresponding to its Bragg wavelength. Small perturbations such as strain or acoustic waves induce shifts in this wavelength, which are transduced into measurable intensity variations when the laser is positioned on the slope of the notch. Precise localization of this spectral feature is therefore critical for high-sensitivity sensing.

To achieve this, the ML-enabled tuning framework is used to perform fine, continuous wavelength sweeps across the FBG spectrum. As shown in Fig. 5b, the laser performs a high-resolution mode-hop-free scan, generating smooth and repeatable spectral traces. The supervised algorithm identifies the notch position by analyzing the power response and detecting the minimum in the reflection profile. This process is repeated for all FBGs distributed across multiple plates, enabling each laser to autonomously discover and map the spectral locations of the sensors.

The resulting DAQ traces in Fig. 5c demonstrate consistent identification of the notch across all lasers and plates, with clear intensity dips corresponding to the FBG resonances. Importantly, the fine-tuning capability ensures that the laser remains continuously aligned to the notch region, even in the presence of temperature-induced drift. Since both the cavity modes and the FBG Bragg wavelength shift with temperature, coarse tuning alone would lead to misalignment and loss of sensitivity. In contrast, the ML-based fine tuning dynamically compensates for these variations by continuously re-optimizing the laser operating point, maintaining stable lock to the notch. This enables robust, repeatable sensing performance under varying environmental conditions, which is essential for practical deployment in real-world structural monitoring applications.

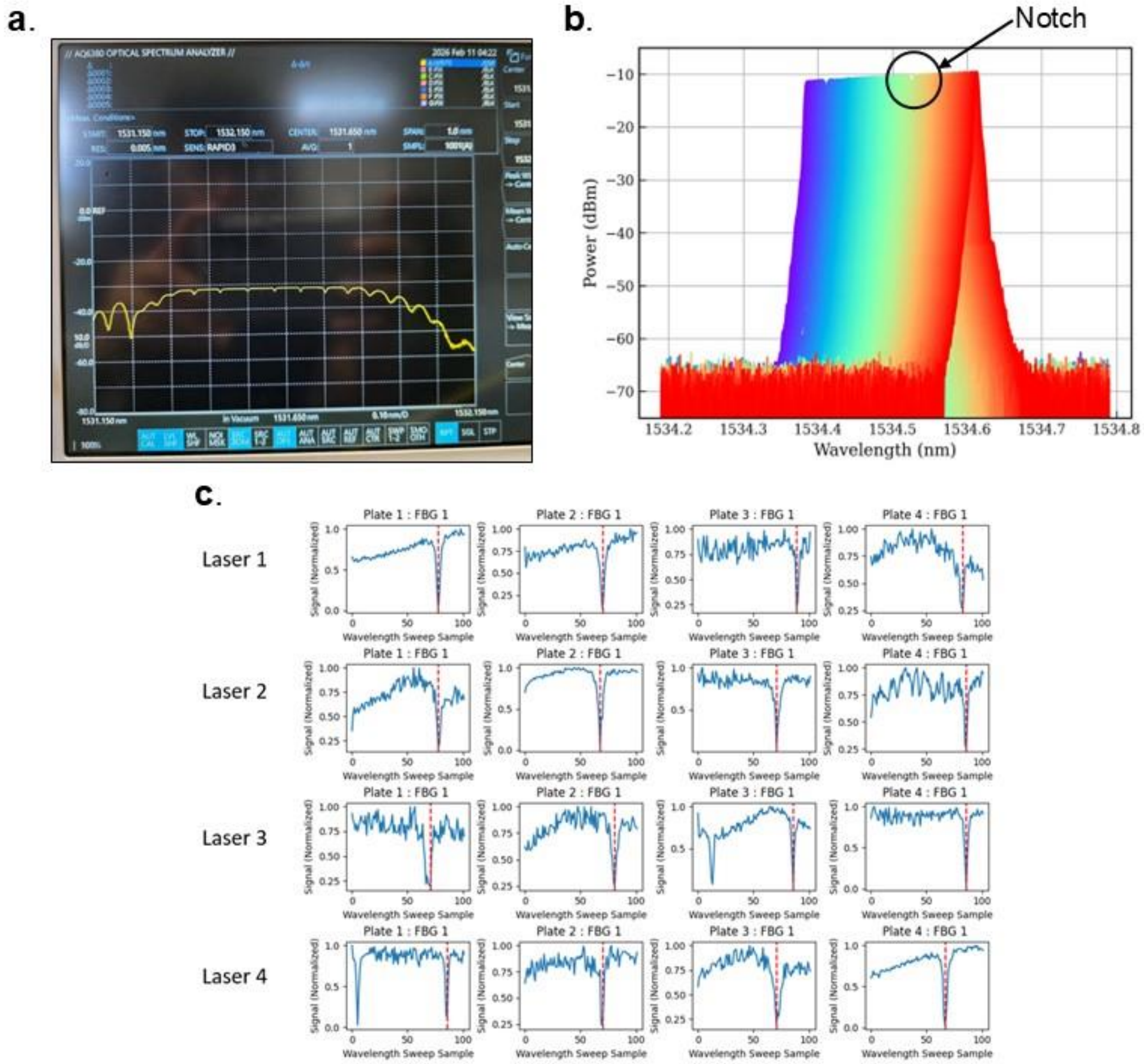

**Fig. 5 | Identification and tracking of FBG reflection notches using ML-trained tunable lasers. a** Measured optical reflection spectrum of a representative FBG, showing the characteristic reflection notch within the broadband optical response. **b** Superimposed spectra obtained during fine, mode-hop-free wavelength sweeps of the tunable laser across the resonance of one of the FBG. The continuous sweep enables precise identification of the reflection notch, highlighted, with high spectral resolution and minimal power variation. **c** DAQ traces of normalized photodetector signals for four lasers sweeping across FBGs distributed on multiple plates. Each trace exhibits a distinct intensity dip corresponding to the FBG reflection notch (indicated by dashed lines), demonstrating the ability of all lasers to accurately locate and lock onto sensor resonances across the distributed network.

To validate the sensing performance of the ML-enabled FBG interrogation system, controlled AE events were generated using PLB tests, which are widely used as repeatable and standardized sources of broadband transient acoustic signals. The PLB event mimics micro-crack formation by producing a sharp, localized release of elastic energy that propagates through the structure and is captured by nearby sensors.

The system works in two different modes before and after detection of AE. Initially, in the first mode of operation, the system is configured such that each of the four lasers is aligned to an FBG located on a different plate. As shown in Fig. 6a, the resulting energy ratio signals exhibit distinct peaks occurring at different time instances, corresponding to AE events originating from different spatial locations. The energy ratio is computed using a short-term average to long-term average (STA/LTA) detection algorithm [20], where the ratio of signal energy in a short time window to that in a longer baseline window highlights transient acoustic events as sharp peaks above the background noise. This demonstrates the distributed sensing capability of the system, where multiple regions can be monitored simultaneously and independently. The variation in signal amplitude and temporal response across plates arises from differences in propagation paths, attenuation, and coupling of the acoustic wave to the respective FBG locations. In a second mode of operation, all four lasers are dynamically reconfigured to lock onto FBGs located on the same plate. Upon occurrence of a PLB event on that plate, all channels detect the acoustic signal nearly simultaneously, as shown in Fig. 6b. The consistency in timing across the channels confirms the ability of the system to localize events and concentrate sensing resources on regions of interest. Minor variations in amplitude are attributed to differences in FBG positioning and local strain sensitivity. Furthermore, the system demonstrates the ability to autonomously transition between these two modes. When an AE event is detected at any location, the ML-controlled lasers reallocate their interrogation wavelengths to neighboring FBGs, enabling enhanced spatial resolution and improved signal fidelity for subsequent events. This adaptive sensing strategy, combined with high-resolution, mode-hop-free laser tuning, enables robust detection of transient acoustic signals across distributed structures. The strain sensitivity measurement shown the sensitivity of these detections can reach up to 358 $f\varepsilon/\sqrt{Hz}$ (See SI-II). Overall, the PLB experiments confirm that the proposed system can reliably detect, localize, and track acoustic emission events in real time, highlighting its potential for scalable structural health monitoring applications. A method for physical deployment of these FBGs using paint is demonstrated in SI-III.

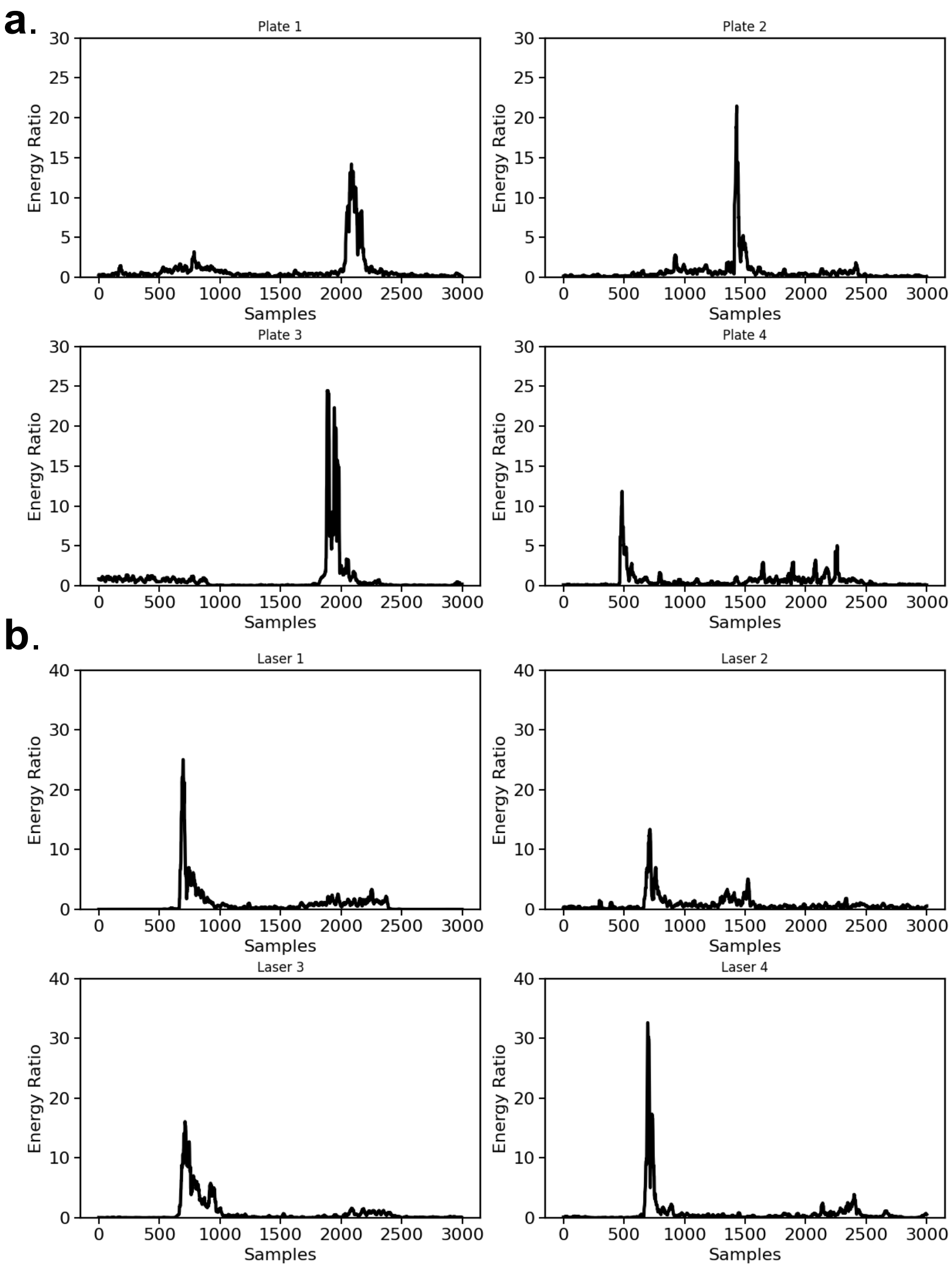

**Fig. 6 | Multi-laser detection of acoustic emission (AE) events across distributed FBG sensors. a** Energy ratio signals recorded by four independent lasers monitoring FBGs located on four different plates. Each laser detects acoustic emission events at distinct time instances corresponding to localized perturbations on separate plates, demonstrating distributed sensing capability. **b** Energy ratio signals when all four lasers are dynamically reconfigured to lock onto FBGs on the same plate. All lasers simultaneously detect the same acoustic event with consistent temporal signatures, confirming coordinated sensing and the ability to localize and track AE events using multiple channels. The DAQ is set at 2.5 kHz acquisition rate and samples on horizontal axis represents the time step corresponding to this acquisition rate.

# Discussion

We have demonstrated an intelligent, distributed acoustic sensing system enabled by machine learning (ML)-trained hybrid III–V/SiN tunable lasers, achieving sub-picometer, mode-hop-free wavelength control and robust multi-channel interrogation of FBG sensors. The advantages of using hybrid III–V/SiN tunable lasers is shown SI-IV. The key advance lies in embedding adaptive intelligence within the laser tuning process, allowing each device to learn and compensate for its own nonlinear electro-thermal characteristics, fabrication variations, and environmental perturbations. This capability addresses a longstanding limitation of external-cavity tunable lasers, where precise and continuous wavelength control is difficult to achieve using conventional open-loop or lookup-based approaches.

The supervised ML framework enables reliable alignment of the laser wavelength to the slope of FBG reflection notches with high precision and stability. This is critical for acoustic sensing, where small wavelength shifts induced by strain must be accurately transduced into intensity variations. The demonstrated sub-picometer tuning resolution (<0.1 pm) and low power variation (<0.5 dB) allow consistent operation at optimal sensitivity points, even in the presence of temperature drift and thermal crosstalk. Importantly, the ability to dynamically re-optimize the tuning parameters ensures that both laser cavity modes and FBG resonances remain aligned over time, significantly improving robustness compared to static calibration methods.

The distributed sensing architecture further highlights the advantages of combining tunable photonic sources with FBG networks. By multiplexing multiple ML-trained lasers, the system enables simultaneous interrogation of spatially separated sensors and adaptive reconfiguration in response to detected events. PLB experiments demonstrate both distributed monitoring, where independent events are detected across different plates; and regional tracking, where multiple lasers converge on a single region of interest. This dynamic allocation of sensing resources represents a significant step toward intelligent structural health monitoring systems capable of prioritizing critical regions in real time.

Compared to distributed sensing techniques based on Rayleigh, Brillouin, or Raman scattering, the FBG-based approach offers higher signal-to-noise ratio, precise localization, and compatibility with high-frequency (>10 kHz to MHz) acoustic detection. At the same time, the use of integrated, narrow-linewidth, widely tunable lasers reduces system complexity by eliminating the need for bulky interrogation units or external scanning mechanisms. The demonstrated architecture is inherently scalable, as additional lasers and sensors can be incorporated to increase spatial coverage and resolution.

Despite these advances, several challenges remain. The current implementation relies on external measurement and control hardware, and further integration of the ML control loop into embedded electronics or on-chip systems would be beneficial for reducing size, weight, and power (SWaP). Additionally, while the supervised learning approach is effective for calibration and tuning, extending the framework to incorporate predictive or adaptive models for long-term drift compensation and fault detection could further enhance system reliability. Finally, real-world deployment will require validation under more complex and dynamic environmental conditions, including varying temperature gradients, mechanical loading, and noise sources.

In summary, this work establishes a new paradigm for intelligent photonic sensing systems, where machine learning is directly integrated into the physical operation of tunable lasers. The combination of high-resolution wavelength control, distributed sensing capability, and adaptive system behavior opens new opportunities for real-time monitoring of critical infrastructure, aerospace systems, and industrial assets.

## Acknowledgements

This material is based upon work supported by the Office of Naval Research under contract N6833522C0834.

# Supplementary Information

## Table of Contents

# I. FBG Spectra

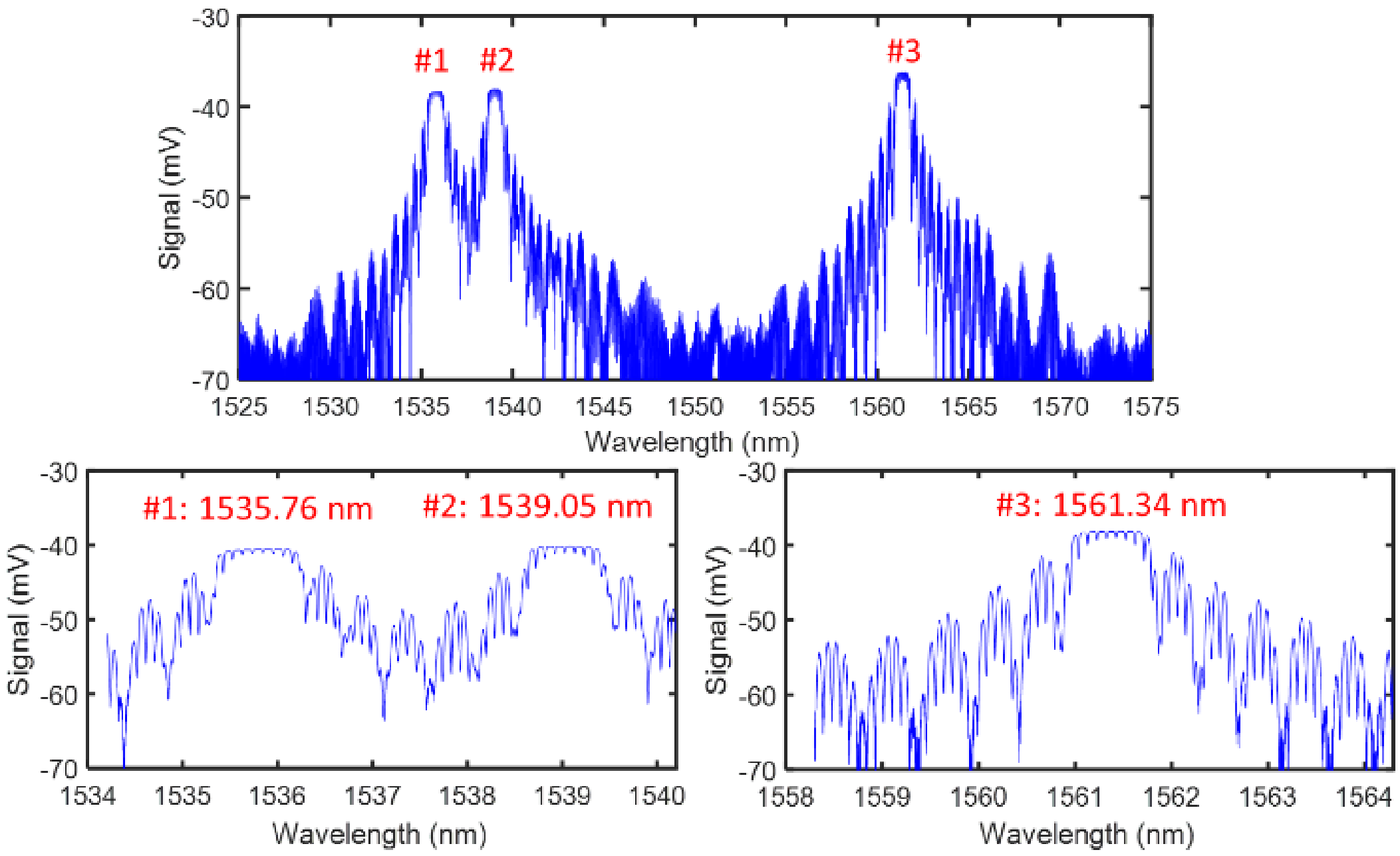

**Fig. S1 | Representative fiber Bragg grating (FBG) reflection spectra at different wavelengths.** Top: Broadband optical spectrum showing multiple FBG reflection notches distributed across the wavelength range, labeled #1, #2, and #3. Bottom: Zoomed-in views of individual FBG notches located at 1535.76 nm (#1), 1539.05 nm (#2), and 1561.34 nm (#3), respectively. Each FBG exhibits a distinct narrowband reflection feature with a characteristic spectral profile, demonstrating wavelength multiplexing capability. These results highlight the ability to distribute multiple FBG sensors across a wide spectral band for simultaneous interrogation using tunable laser sources.

# II. Strain sensitivity

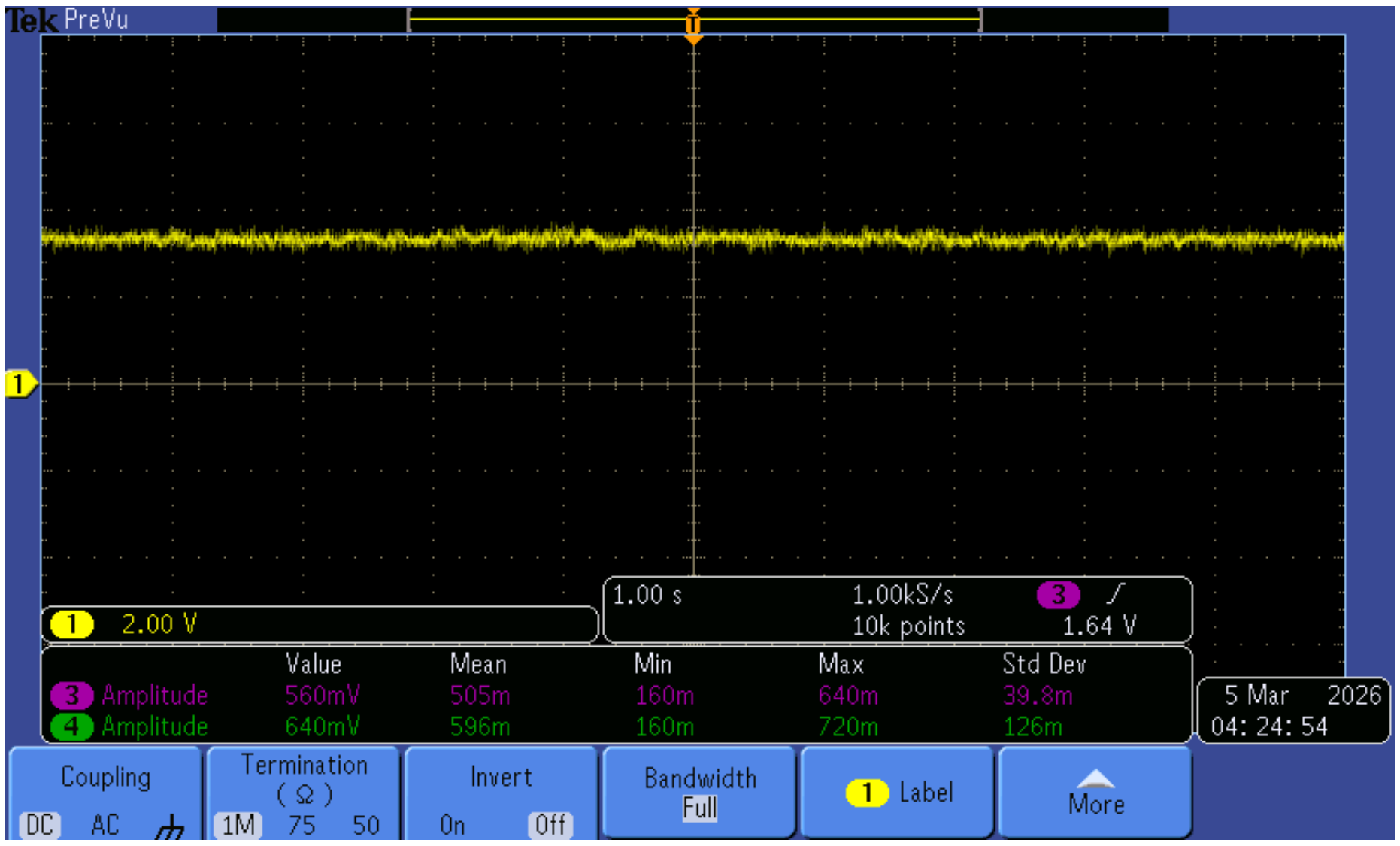

**Fig. S2 | Sensitivity measurement of the FBG interrogation system.** Oscilloscope trace showing the measured voltage fluctuations used to estimate the strain sensitivity of the system. The recorded AC noise ($V_{rms}$) and DC signal levels are used to compute the wavelength noise equivalent and corresponding strain. Using the measured parameters and system bandwidth, the resulting strain sensitivity is calculated to be 358 fε/√Hz, demonstrating high-resolution detection capability suitable for acoustic emission sensing.

## III. Painted plate installation

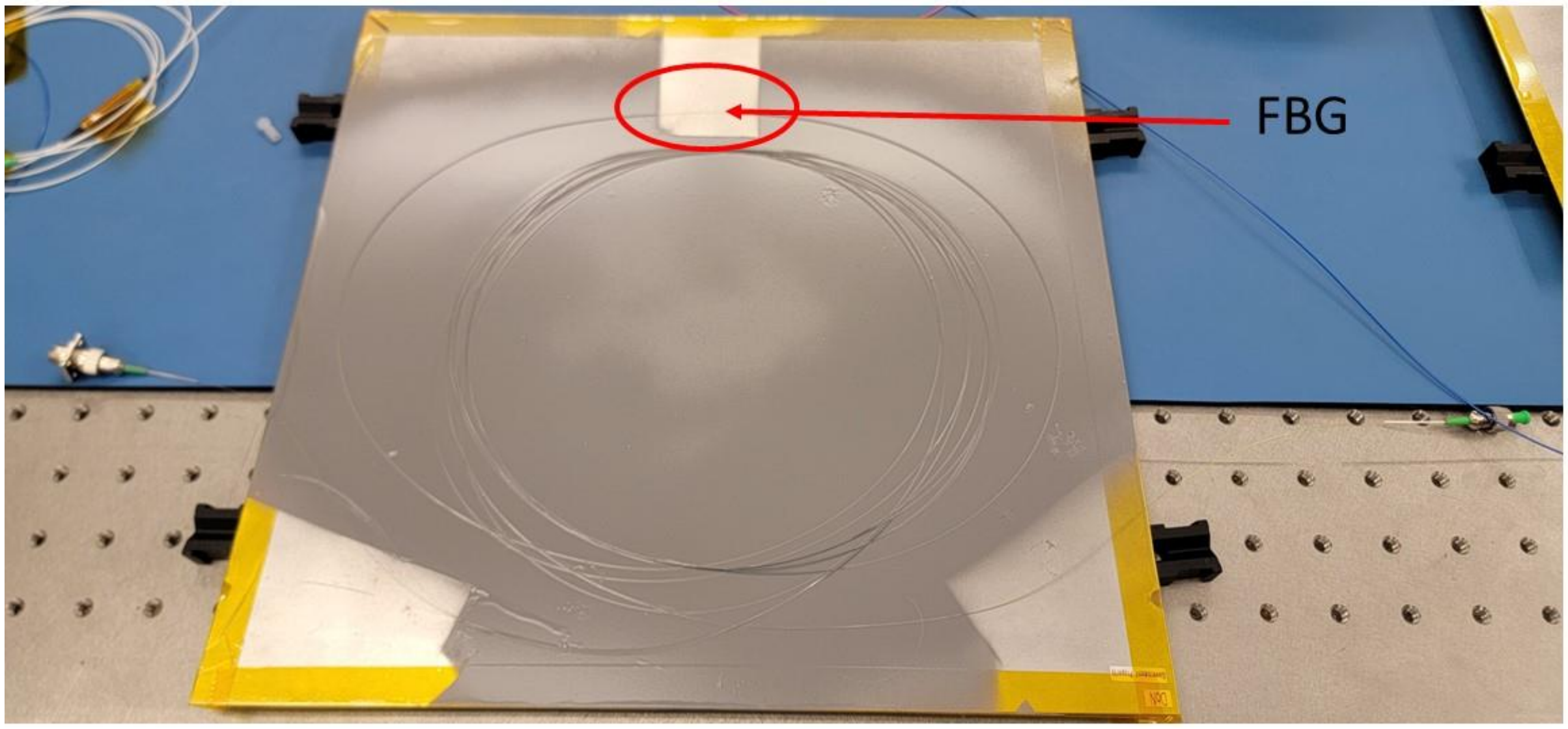

**Fig. S4 | Practical deployment of FBG sensors on coated structural surfaces.** Photograph of an aluminum plate with an optical fiber containing fiber Bragg gratings (FBGs) bonded onto the surface and covered with a protective coating. The FBG location is highlighted. The coating layer secures the fiber to the structure while providing mechanical protection and improved coupling of acoustic strain from the substrate to the sensor. This approach enables robust and practical deployment of FBG-based sensing systems for real-world structural health monitoring applications.

## IV. Comparison table

| Feature | ISET Laser (Integrated Silicon External-Cavity Tunable) | External-Cavity Laser (ECL) | DFB Laser (generally non-tunable) |
|---|---|---|---|
| **Alignment** | Edge or vertical integration | Free-space / fiber coupling | Monolithic |
| **Tunability** | Yes (Si-based tunable filters, wide range) | Yes (wide range, mechanical tuning) | Limited (small electrical tuning range) |
| **Linewidth / RIN** | Narrow linewidth, low RIN | Narrow linewidth, low RIN | Limited linewidth control, higher RIN |
| **Mode-hop Stabilization** | Yes (ML-enabled – This work) | Limited (requires complex control) | Not applicable |
| **Size / SWaP** | Compact, low SWaP | Bulky, high SWaP | Compact, low SWaP |
| **Cost** | Moderate (scalable integration) | High | Low |

**Table S1 | Comparison between ISET laser vs ECL vs DFB Laser for tunable lasers.**